\documentclass[11pt]{article}
\usepackage{epsfig,endnotes}
\usepackage{algorithm}
\usepackage{algorithmic}
\usepackage[small,compact]{titlesec}
\usepackage{url} 
\begin{document}



\title{How to Bypass Verified Boot Security in Chromium OS}
\author{
{\rm M. I. Husain}\\
University at Buffalo
\and
{\rm L. Mandvekar}\\
University at Buffalo
\and
{\rm C. Qiao}\\
University at Buffalo
\and
{\rm R. Sridhar}\\
University at Buffalo
} 

\maketitle

\thispagestyle{empty}

\begin{abstract}
Verified boot is an interesting feature of Chromium OS that supposedly can detect any modification in the root file system (rootfs) by a dedicated adversary. However, by exploiting a design flaw in verified boot, we show that an adversary can replace the original rootfs by a malicious rootfs containing exploits such as a spyware or keylogger and still pass the verified boot process. The exploit is based on the fact that a dedicated adversary can replace the rootfs and the corresponding verification information in the bootloader. We experimentally demonstrate an attack using both the base and developer version of Chromium OS in which the adversary installs a spyware in the target system to send cached user data to the attacker machine in plain text which are otherwise encrypted, and thus inaccessible. We also demonstrate techniques to mitigate this vulnerability.
\end{abstract}



\section{Introduction}
\label{introduction}
Chromium OS~\cite{cros} is a web-centric operating system with Chromium browser as its principal user interface. It is designed for the users who often spend considerable amount of time on the Internet. Since the system is primarily optimized with components to support the browser, it can boot really fast ($~10$ seconds).

The Chromium OS design document~\cite{secoverview} discusses the following use cases for Chromium devices: (i) ubiquitous computing, (ii) use as a secondary entertainment computer, (iii) lending device to the customers of coffee shops and libraries, and (iv) sharing a common computer among family members.

Since sharing a device among multiple users is one of the intended use cases, Chromium OS is designed to support scenarios such that one user (even the owner) cannot access the data of another user. Each user logs in to the system using his/her Google credentials (username and password). Once logged in, the users can access their applications as web applications on the Chromium browser. When the user logs out, his/her cached data~\cite{pcud} is encrypted using his/her credentials. Therefore, it is not possible for a user to access another user's data without decrypting it with the original user's credentials. All these activities take place at the front-end of the system (Google account user domain) which is not tightly coupled with the core of the system (kernel and rootfs). This design decision is to facilitate the update or recovery of the core of the system without affecting the user settings and data. More details on cached user data security is in Section~\ref{pcud}.


On top of that, Chromium OS employs a technique called verified boot~\cite{verifiedboot} that can check the system integrity between two boot sequences. This feature verifies the integrity of the rootfs to make sure that there was no intended or unintended modification of it. Such modifications include but are not limited to adding a malicious superuser to the rootfs and installation of a rootkit or malware. Verification of the rootfs is very important from security point of view, because as mentioned earlier, the front-end of the system (Google account user domain) is not tightly coupled with the rootfs. In other words, modification to the rootfs does not directly affect the user settings in the front-end. Without any verification mechanism in place, modifications to the rootfs can go undetected and users can be vulnerable to the modifications. More on the details of the verified boot process is in Section~\ref{verified_boot}.

To identify such modifications, Chromium OS stores the verification information of the original rootfs with the bootloader and uses it to verify the integrity of the rootfs during the next boot process. In the disk layout, Chromium OS has 12 partitions (more details in Section~\ref{disk_format}). Partition 1 is used to store encrypted user data as discussed earlier. Partition 3 contains the rootfs. Partition 12 contains the kernel command line as well as rootfs verification  information. In the current verified boot process, the rootfs verification  information stored in partition 12 is compared with the checksum of the rootfs at partition 3. Also, the verified boot feature disables external mounting of the rootfs partition (3) and the partition is read-only for added security.

In this paper, \textit{we demonstrate techniques that a dedicated adversary can exploit to take complete control over the rootfs bypassing the verified boot process}. In other words, the adversary can replace the original rootfs with his/her malicious rootfs using these techniques, but the rootfs verification process will be unable to detect it. To realize these exploits, key challenges that our proposed techniques overcome are: (i) how to enable the mount option of partition 3 and (ii) how to gain read-write access to the partition. We briefly summarize our techniques below:

\begin{enumerate}
\item \emph{Exploit 1}: Since the rootfs verification hash in partition 12 is in unencrypted form and can be manually modified, an adversary can modify the  information to match any malicious rootfs replacing the original rootfs at partition 3. More specifically, first the adversary builds his/her own Chromium OS image and modifies the rootfs to contain a malware. After that, the adversary overwrites the victim's partition 3 and partition 12 using bit-by-bit copy~\cite{dd} with his/her own (malicious) partition 3 and partition 12. Using bit-by-bit copy, the adversary can overcome the challenge of mounting the partition and gaining read-write access. 
\item \emph{Exploit 2}: An adversary with advanced technical know-how can edit the victim's rootfs partition using a hex editor~\cite{hexeditor} to remove the information that blocks the mounting and read-write capability. Once this is done, the adversary can mount the rootfs, remove verification feature and install malware as with exploit 1. 
\end{enumerate}

\textit{How can the adversary extract useful information, such as encrypted cached user data from the front-end (partition 1)}? Once the adversary has complete control over the rootfs, it can install a malware that will snoop a user's data once he/she logs in using Google credentials (note that at this point, that particular user's data is decrypted and exists in plaintext on partition 1) and send it to a machine of adversary's choice. Although we only demonstrate this simple spyware for proof-of-concept, an adversary can practically install much severe malware such as a keylogger to steal a wide range of sensitive user information including online credentials.
\begin{figure*}[htbp]
  \centering
    \includegraphics[scale=0.3]{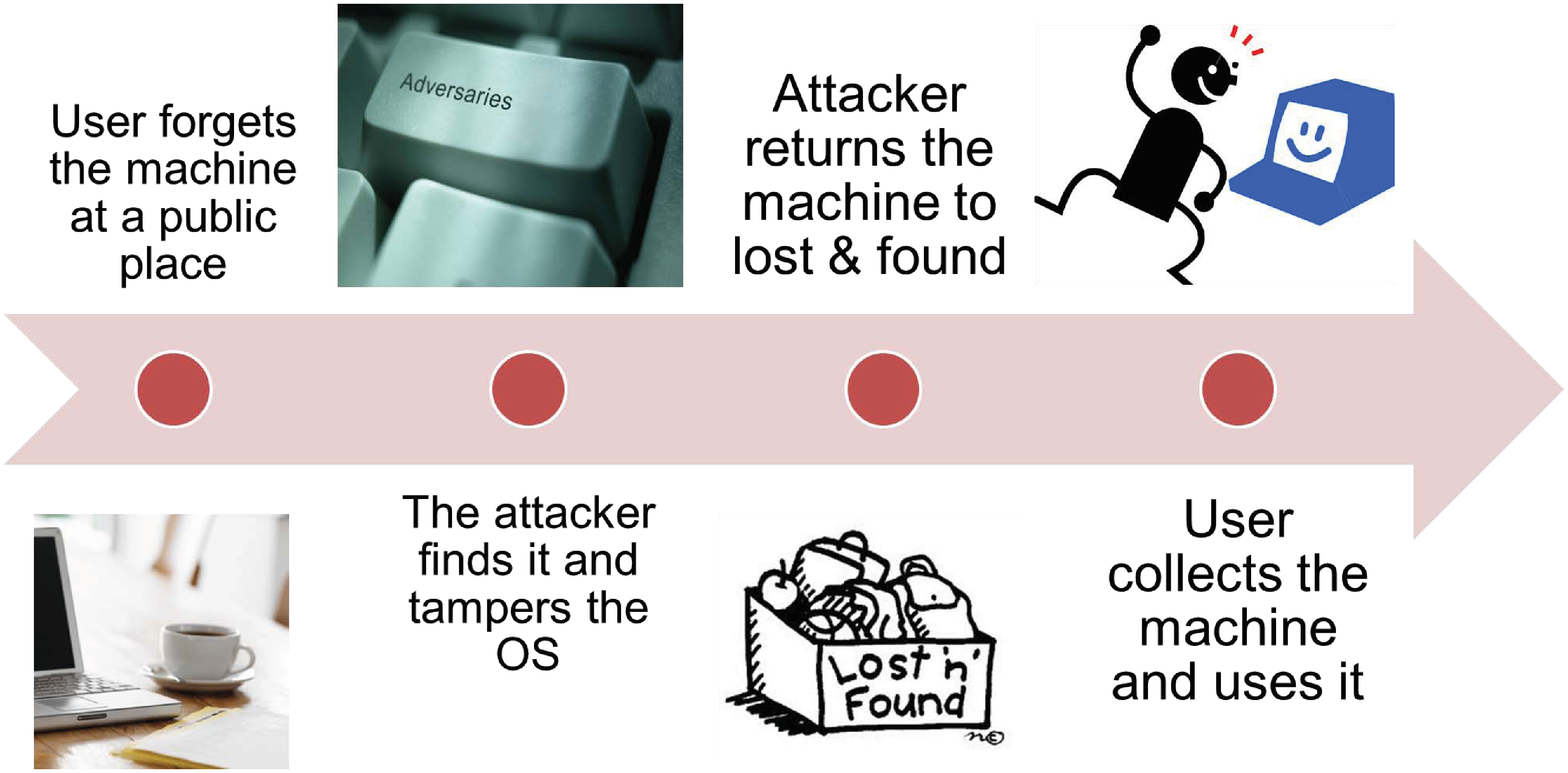}
\caption{\small{Example scenario of the demonstrated attack}}
\label{fig:attack-example}
\end{figure*}

\textit{How can this attack take place in a real usage scenario?} Say, a user, Alice, uses Chromium OS with verified boot support on her netbook~\cite{netbook}. One day, she forgets her netbook in the library (Figure~\ref{fig:attack-example}). Attacker Eve finds the netbook, takes the disk out, overwrites the original rootfs partition with her tampered rootfs partition containing a spyware or keylogger. After that, Eve puts the disk back and returns the netbook to the lost-and-found section of the library. Alice picks up the netbook from lost-and-found, and boots up the machine. Alice believes that if there is any modification of the system, the verified boot will alert her. Since her netbook passes a smooth boot process, Alice thinks that her system is intact and happily starts using the machine. As soon as Alice starts using the machine, the spyware installed by Eve is invoked and it starts covertly sending Alice's sensitive information to Eve. 


On the flip side, another potential application of our technique could be in \textit{digital forensic investigation}. There were several recent incidents~\cite{fde} where law enforcement agencies had confiscated the computing/mobile device of the suspect, but couldn't access it since the suspect was not providing the password (In the United States, the suspect can plea the Fifth Amendment which protects against self-incrimination, including disclosure of encryption keys in some cases). In such an event, where the suspect device is a Chromium device, forensic investigators can acquire a court order~\cite{wiretap} to eavesdrop on the suspect's communication using our proposed method. 

Apart from demonstrating both the techniques, we also discuss how to mitigate such scenarios. Our goal here is to notify the user about the rootfs modification \textit{before} he/she logs in to the system using Google credentials. Also, the mitigation procedure should be fast enough to keep up with boot time performance of Chromium OS. The general flow of our mitigation approach is the following: (i) the verification information of the rootfs is stored in partition 1 in encrypted form and (ii) during the next boot, the user invokes the verification script with decryption password which decrypts the verification information and compares it with the information of the current rootfs. This approach addresses both exploits discussed before. Because, in the first exploit, where the original rootfs is completely overwritten, the absence of verification script functionality notifies the user of the modification. In the second exploit, the script performs the verification as designed. Also, since the information is encrypted using a user supplied password, it can not be easily manipulated. Further, all these verifications take place before the user logs in using Google credentials. So, the per user cached data remains secure in encrypted form in partition 1 while the verification takes place.

To summarize our contributions:

\begin{itemize}
\item We have demonstrated techniques that an adversary can utilize to bypass the verified boot security and steal encrypted user information. These techniques have been experimentally tested on the all available versions of Chromium OS (rootfs-verification-enabled base and developer versions). On the experimental platform, exploit 1 took $10.3$ minutes whereas exploit 2 only took $53$ seconds to bypass the verified boot on an average.

\item These exploits do not assume any knowledge about anyone on the user (victim) machine and certainly do not require prior superuser privilege on the user (victim) operating system with verified boot support. These are also not specific to any hardware configuration and do not make any assumption on the hardware of the user machine. Further, these exploits do not disappear by simply rebooting the system. Note that, many of the attacks described in the Chromium OS verified boot design document are patched by reboot~\cite{verifiedboot}


\item Practical mitigation techniques are proposed and experimentally verified to make any modification of the rootfs evident to the user. On an average, our mitigation techniques took $0.22$ seconds to execute which is very negligible (overall boot time is $~10$ seconds).

\end{itemize}


\section{Chromium OS Overview}

In this section, we briefly discuss different features of Chromium OS that are pertinent to the exploit. 
\subsection{Software Architecture}

Chromium OS has a simple software architecture. The front-end of the system is the Google user domain which contains the browser, window manager and web applications. The front-end is supported by the OS backend (kernel and rootfs) which provides OS functionality and sits on top of the hardware. A unique feature of Chromium OS is that the front-end of the system is not tightly coupled with the backend of the system. For example, per device system settings such as locale, WiFi settings and owner of the system information are managed by the system backend. However, per user settings such as browser setting, account preferences, web applications and autofill data are managed by the front-end of the system (after the user logs in using his/her Google credentials)~\cite{useraccount}. Due to this fact, it is possible to share a Chromium device when protecting the confidentiality of each user's data which we discuss next. However, lack of tight coupling between the Google user domain and rootfs domain also make it difficult to notice any modifications in the underlying core of the system. Therefore, verification of rootfs is an essential security feature for operating systems like Chromium OS. 
\begin{figure}[htbp]
  \centering
    \includegraphics[scale=0.5]{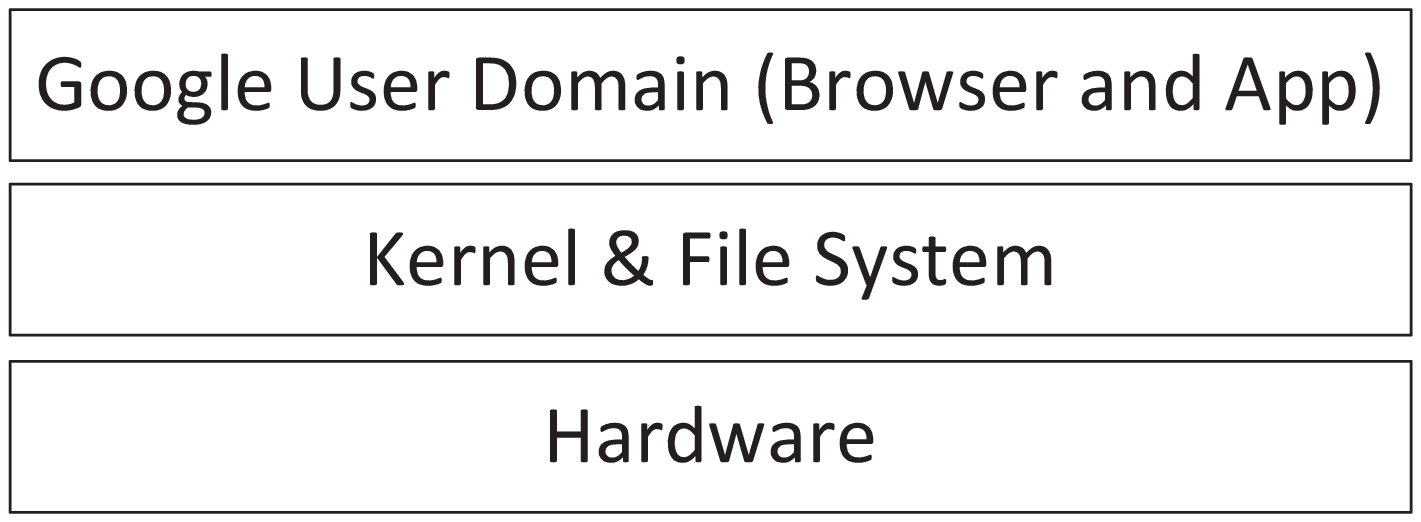}
\caption{\small{Software architecture of Chromium OS}}
\label{fig:software-architecture}
\end{figure}

\subsection{Protecting Cached User Data}
\label{pcud}
Chromium OS allows multiple users to access a given device, but it doesn't allow data belonging to one user to be seen by other
users. Consider a Chromium OS device belonging to Alice. Bob, a friend of Alice, borrows the device from Alice for sometime and performs online activity using his Google credentials. Even though the device is owned by Alice, she has no way to find out what files, or websites were accessed by Bob. Of course, neither can Bob find out any data belonging to Alice.

To support this, Chromium OS encrypts the user's data. More specifically, it encrypts each user's files in the home directory as well as cached browser data (Figure~\ref{fig:verified-image-part1-encrypted-files}). In addition, data created and maintained by plugins and web applications are also encrypted.

In typical operating systems the root or administrative user (owner) can access all the users' files. Therefore, in the front-end (Google user domain) Chromium OS enforces per-user encryption instead of just relying on file ownership and access control to prevent users of a system to access each other's files. Encryption better suits this problem since the root or administrative user would still have to recover the encryption keys to be able to access the other user's data. Technically, a unique vault directory and keyset is assigned to a user at the first login using Google credentials.  The vault is nothing but an encrypted storage for a particular user data.  The keyset is tied to the user's login credentials and is required to allow the system to both retrieve and store information in the vault.  When the user logs in, the vault is open for access by the user. Therefore, \emph{if an attacker can run a malware in the system when the user is logged in, it can have access to all of user's data in decrypted form}. In this paper, we demonstrate that such an attack is possible by modifying the rootfs and bypassing verified boot.

\subsection{Disk Format}
\label{disk_format}
Chromium OS is a customized GNU/Linux distribution. Bootable Chromium OS drives have a common drive format where the partition contents are as shown in Table~\ref{tab:partition-table}. Among these partitions, only partition 1, partition 3 and partition  12 are relevant to our paper.
\begin{table*}[htbp]
  \centering
  \caption{GUID Partition Table of Chromium OS}
  \scalebox{0.65}{
    \begin{tabular}{ccc}
    \hline
    Partition       & Usage  &  Purpose \\
    \hline
     1    &  Cached user data & User's browsing history, downloads, cache, etc. Encrypted per-user \\
     2    &  kernel A &  Initially installed kernel \\
     3    &  rootfs A &  Initially installed rootfs  \\
     4    &  kernel B &  Alternate kernel, for use by automatic upgrades \\
     5    &  rootfs B &  Alternate rootfs, for use by automatic upgrades \\
     6    &  kernel C &  Minimal-size partition for future third kernel \\
     7    &  rootfs C &  Minimal-size partition for future third rootfs. Same reasons as above \\
     8    &  OEM customization &  Web pages, links, themes, etc. from OEM \\
     9    &  reserved &  Minimal-size partition, for unknown future use \\
     10   &  reserved &  Minimal-size partition, for unknown future use \\
     11   &  reserved &  Minimal-size partition, for unknown future use \\
     12   &  EFI System Partition &  Bootloader and rootfs verification info \\
    \hline
    
    \end{tabular}
    }
  \label{tab:partition-table}
\end{table*}

\begin{figure*}[htbp]
  \centering
    \includegraphics[scale=0.25]{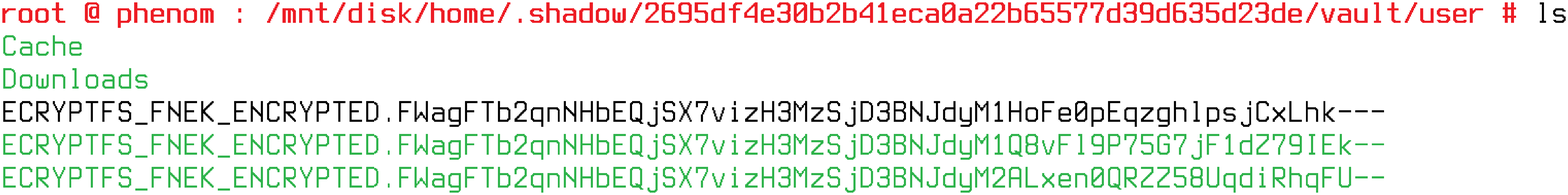}
\caption{\small{Encrypted cached user data on partition 1}}
\label{fig:verified-image-part1-encrypted-files}
\end{figure*}

\begin{figure}[htbp]
  \centering
    \includegraphics[scale=0.3]{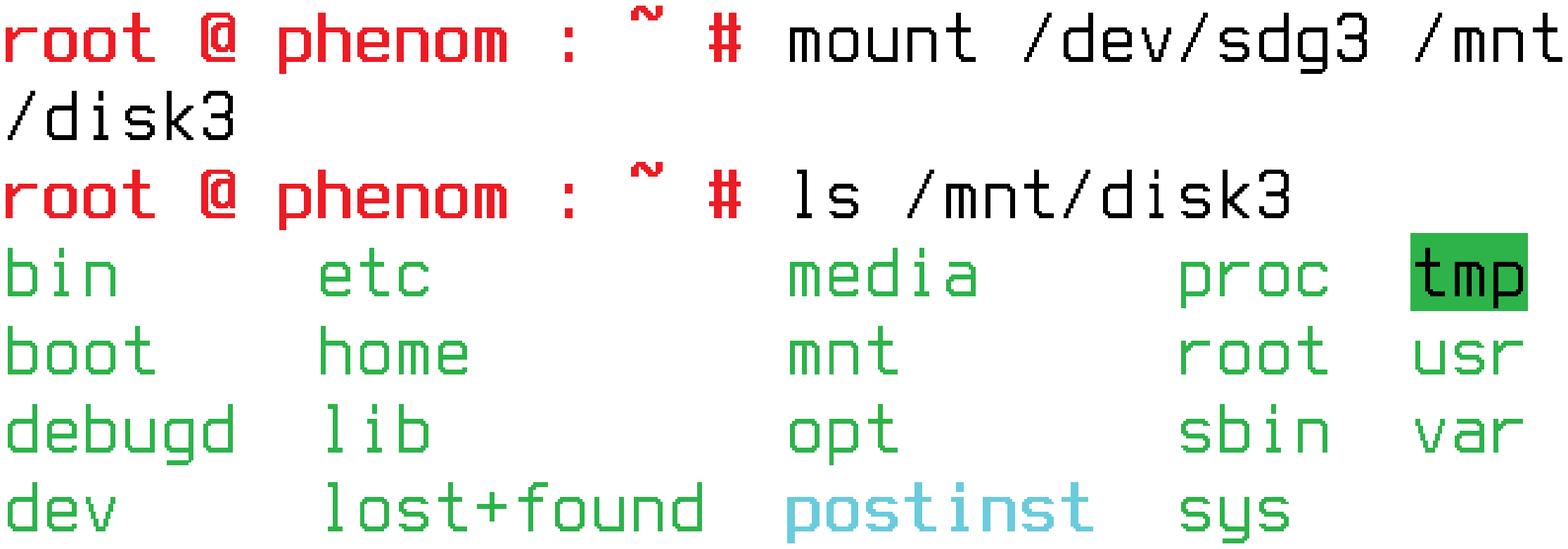}
\caption{\small{rootfs on Partition 3}}
\label{fig:part3-noverify-success}
\end{figure}

Partition 1 mainly contains the user data in encrypted form as shown in Figure~\ref{fig:verified-image-part1-encrypted-files}. Partition 3 contains the rootfs (Figure~\ref{fig:part3-noverify-success}). 

\subsection{Verified Boot}
\label{verified_boot}
\begin{figure}[htbp]
  \centering
    \includegraphics[scale=0.4]{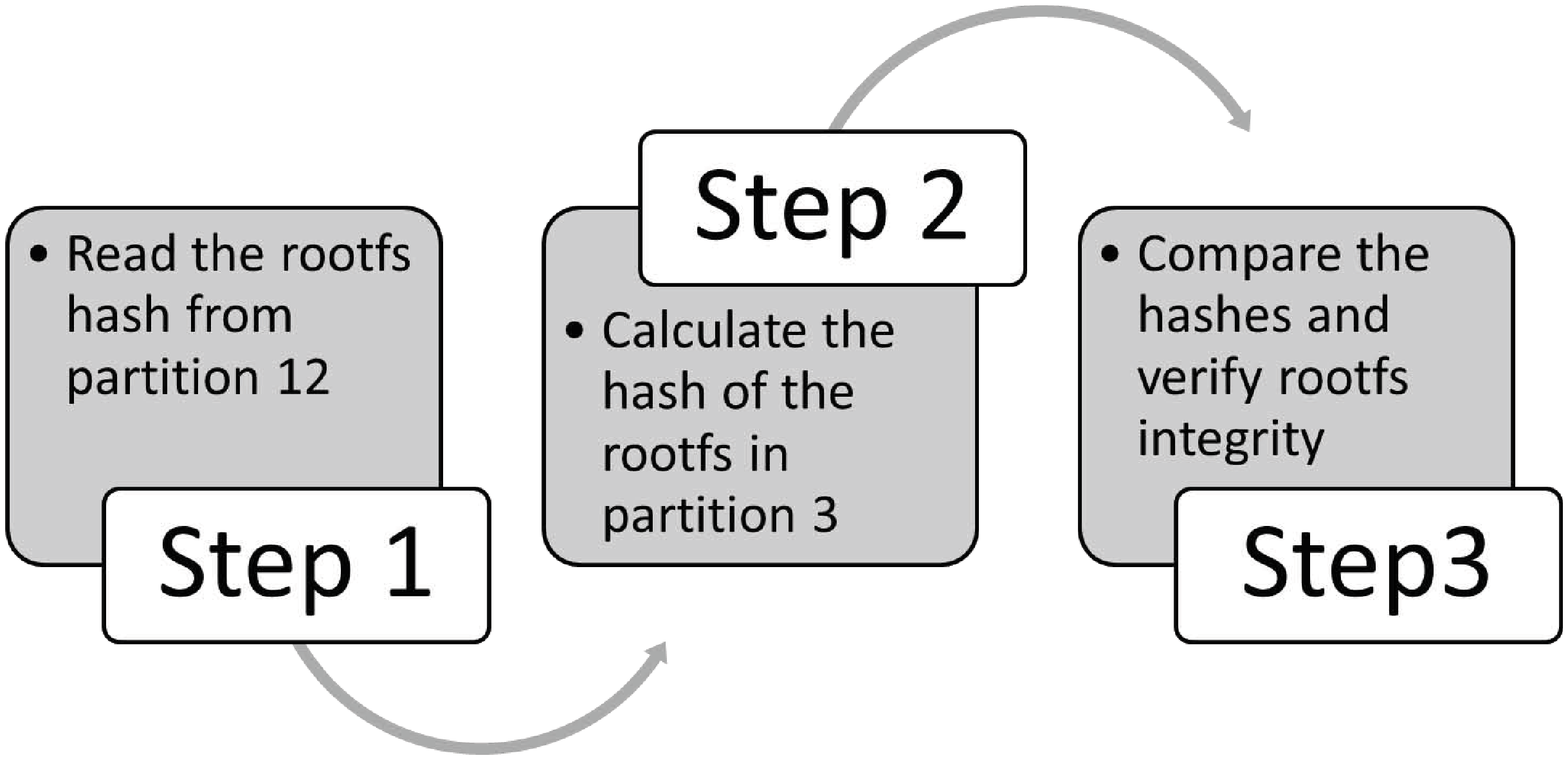}
\caption{\small{Different stages of verified boot}}
\label{fig:verified-boot}
\end{figure}
\begin{figure*}[htbp]
  \centering
    \includegraphics[scale=.28]{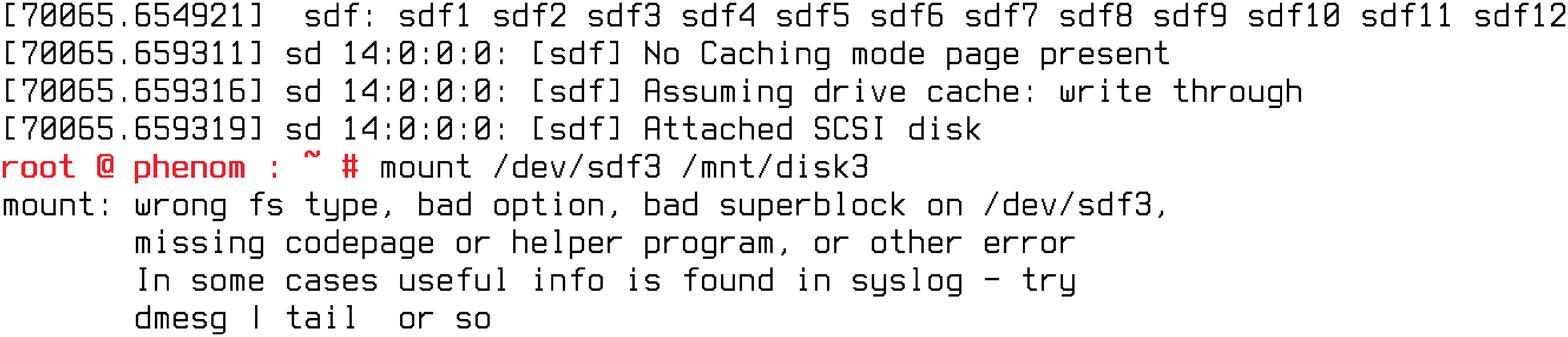}
\caption{\small{Mount failure of partition 3 on Chromium OS with verified boot}}
\label{fig:dmesg-mount-verified}
\end{figure*}

The verified boot in Chromium OS ensures that users feel secure when logging into a Chromium OS device.
While Chromium OS doesn't rule out the possibility of successful attacks,
verified boot is designed in such a way that an attack, once executed,
won't be persistent. When Chromium OS boots, the state of the rootfs is verified and booting is allowed only if the
state is known to be good. Otherwise, a recovery or reinstall procedure
is triggered, which happens outside the usual modes used by attack vectors.
The verified boot is designed to detect any modifications in the file system. Also, for added security, when rootfs verification is enabled, the rootfs partition is read only and not externally mountable (Figure~\ref{fig:dmesg-mount-verified}).
The rootfs verification mechanism takes place in three steps as shown in Figure~\ref{fig:verified-boot}. First, the rootfs verification information such as hash is read from partition 12 (bootloader). Next, the hash of the current rootfs (partition 3) is calculated. If these two hashes match, the system proceeds, otherwise it throws an error (kernel panic) and recovery is initiated. 

\begin{figure}[htbp]
  \centering
   \includegraphics[width=\columnwidth]{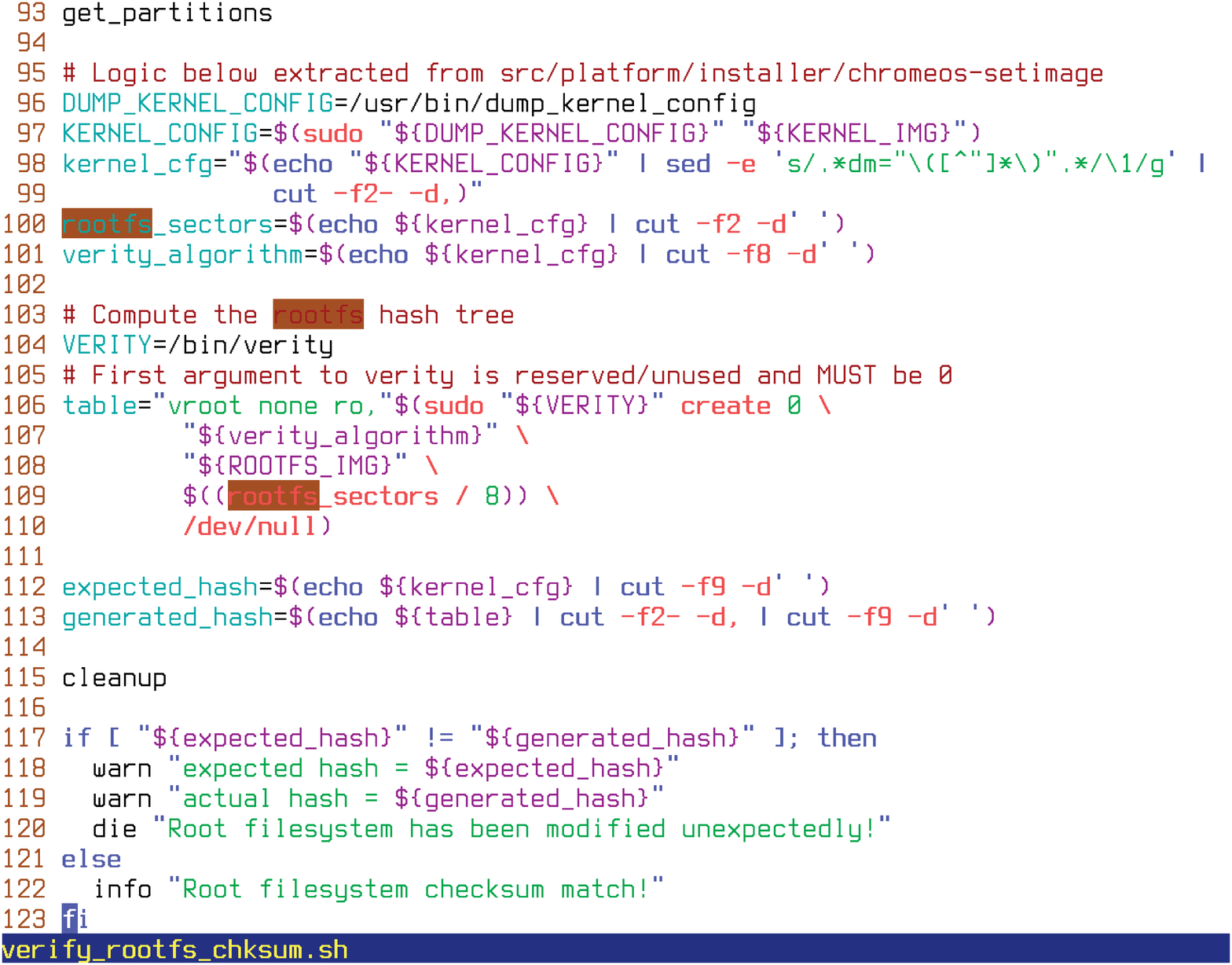}
\caption{\small{Internals of verify rootfs checksum script}}
\label{fig:verify-rootfs-checksum}
\end{figure}

However, the rootfs verification information at partition 12 is stored in unencrypted form and it is possible to overwrite the rootfs verification information to match a malicious rootfs replaced by adversary in partition 3. 

\section{Threat Model}


There are two types of adversaries described by the Chromium OS
threat model:
\begin{itemize}
\item Opportunistic adversary: one who does not target
any specific individual or organization, but will try to gain
access to any vulnerable system available. Such an adversary will
likely sniff network packets to search for vulnerable hosts, or
run websites which could execute malicious code on the vulnerable
system thus giving the adversary control of the system.
\item Dedicated adversary: one who targets only a specific individual
or organization with the explicit purpose of gaining access to or
stealing data from systems belonging to the latter. Such an adversary
is even ready to physically steal the device if needed, in addition to
employing conventional attacks in order to own the vulnerable OS.
\end{itemize}

Although verified boot process can detect if the rootfs has been modified, it doesn't prevent the original rootfs partition from being overwritten by a malicious rootfs partition by a dedicated adversary.

\section{Demonstrating Exploits}
\begin{figure}[htbp]
  \centering
    \includegraphics[width=\columnwidth]{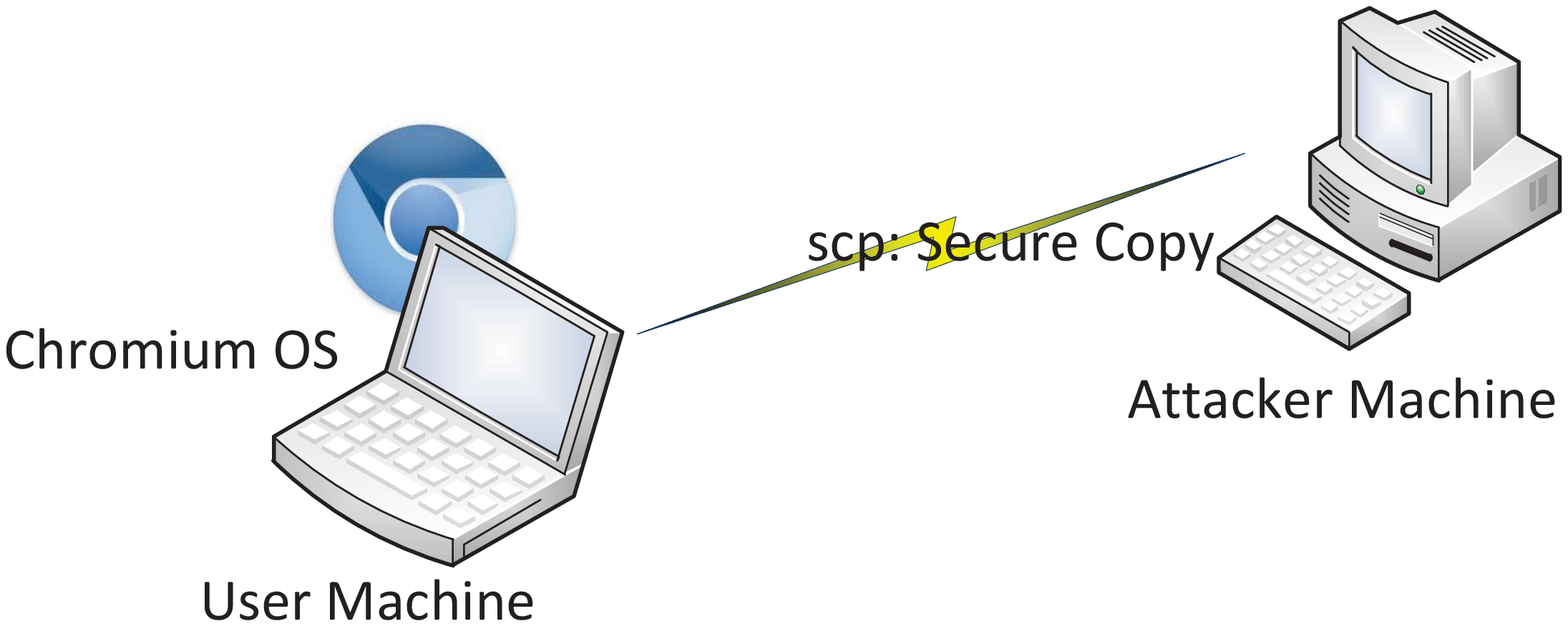}
\caption{\small{Machine setup for the exploit}}
\label{fig:attack-environment}
\end{figure}
For the purpose of demonstrating the exploits, we have built Chromium OS images from the source code~\cite{crossource}. The build environment was a Ubuntu 10.04, 64 bit machine with \texttt{root} access and 12 GB RAM. Chromium OS can be built in two modes: base and developer. The developer image has some additional development packages compared to the base image. The exploits described in this section section apply to both base and developer version of Chromium OS.

As mentioned in the introduction, we demonstrate two techniques to bypass the rootfs verification: The first technique is based on overwriting the original rootfs with a malicious rootfs and the second technique is based on removing the verification functionality from the original rootfs and then installing the malware on it. 

\begin{enumerate}
\item \emph{Overwriting the original rootfs}:The core of the exploit is to overwrite the rootfs partition (partition 3, Table~\ref{tab:partition-table}) of a verified boot Chromium image with a tampered (malicious) rootfs containing a spyware. Therefore, the exploit can be divided into two major parts. First, the attacker builds a Chromium OS image with no verified boot support (rootfs verification disabled) and installs the spyware on it. Next, the adversary physically acquires the disk from the user machine containing Chromium OS with verified boot support and overwrites its rootfs partition with the tampered rootfs prepared in the first phase. We describe these phases in detail in the following sections.

\emph{Tampering the rootfs}: It is possible to disable the rootfs verification during Chromium OS image build process. We exploit this fact to prepare the tampered rootfs partition containing the spyware for the next phase. 
\begin{itemize}
\item After building a Chromium OS image without enabling the verified boot, it was written to a disk. The disk containing Chromium OS and cached user data is mounted to the adversary machine. 
\begin{figure*}[htbp]
  \centering
    \includegraphics[scale=0.2]{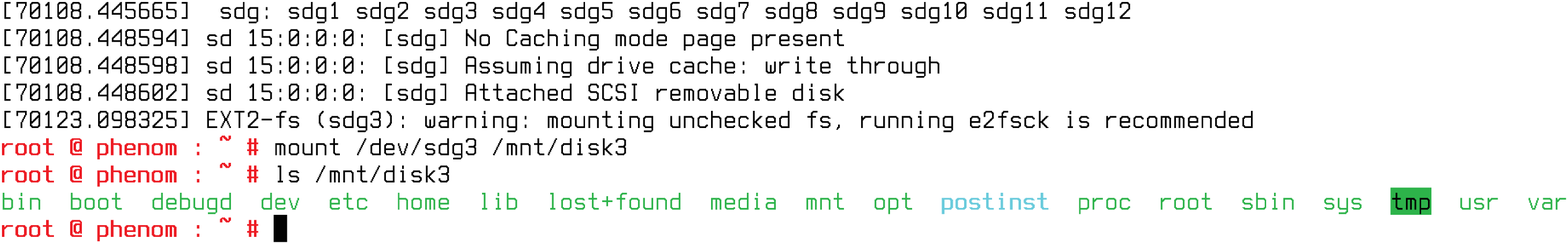}
\caption{\small{Disk with Chromium OS connected as an external drive}}
\label{fig:dmesg-mount-noverify}
\end{figure*}
\begin{figure}[htbp]
  \centering
    \includegraphics[scale=.25]{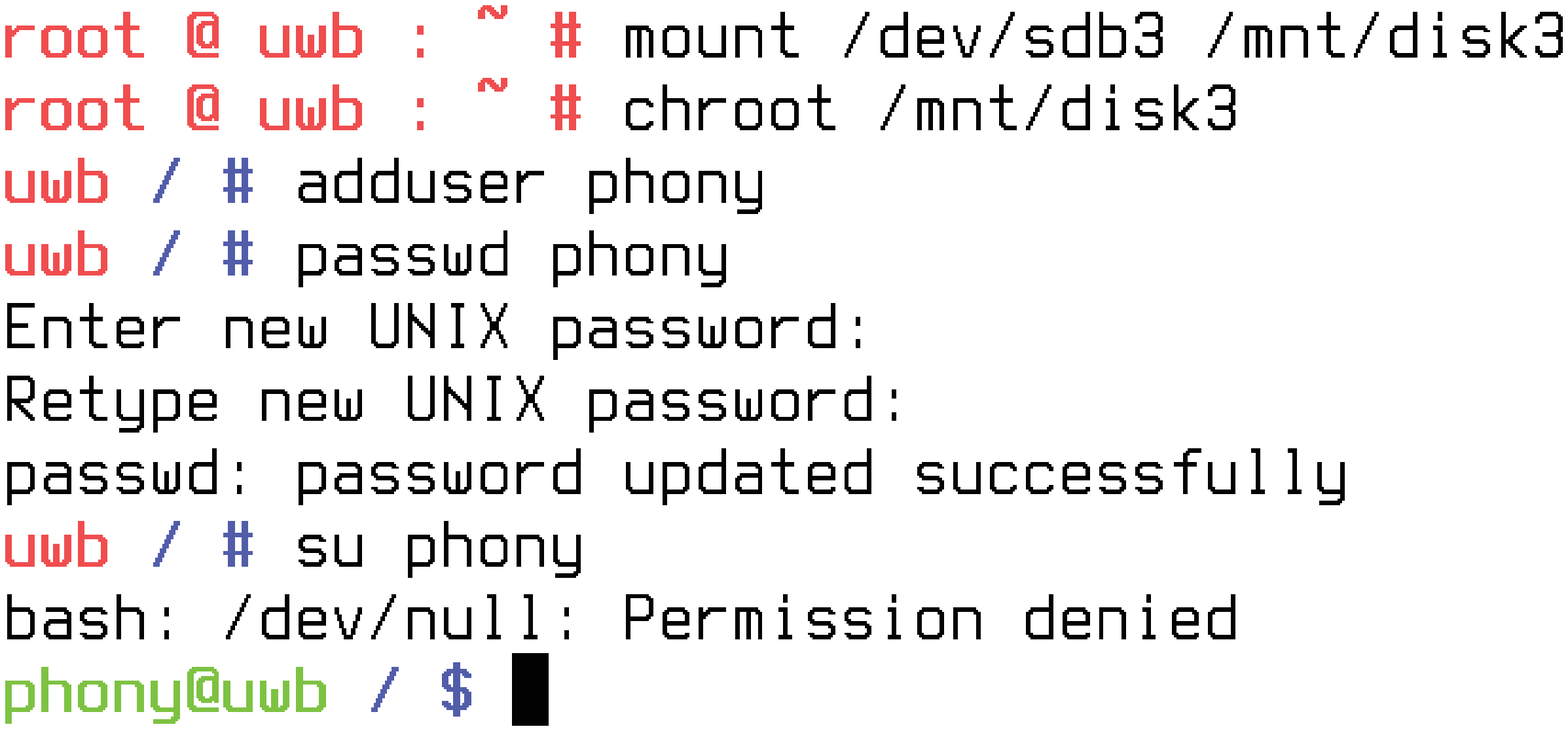}
\caption{\small{Adding a phony user with superuser privilege}}
\label{fig:tampered-image-new-user}
\end{figure}
\item After mounting, the adversary can modify the rootfs (Figure~\ref{fig:part3-noverify-chroot}) \texttt{chroot}ing to the partition and can add a phony user with the superuser privilege (Figure~\ref{fig:tampered-image-new-user}). 
Next, we show how to overwrite this tampered rootfs partition to a Chromium OS image with verified boot support.
\begin{figure}[htbp]
  \centering
    \includegraphics[scale=.25]{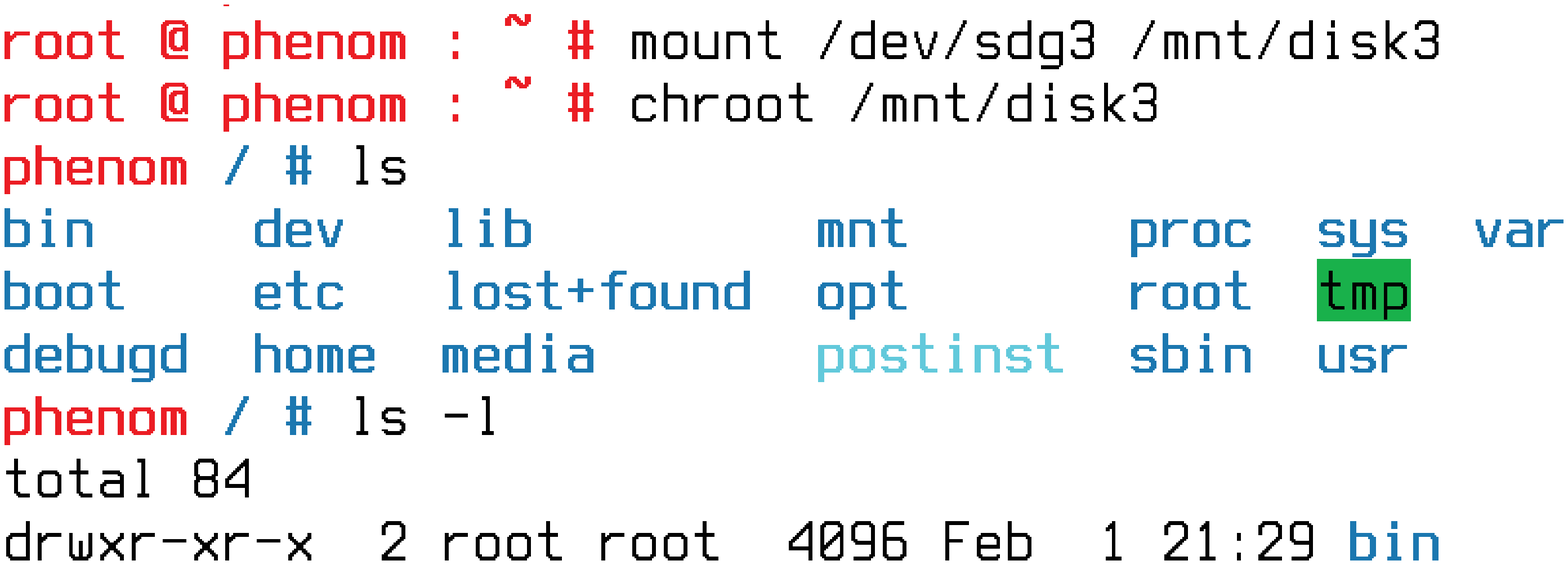}
\caption{\small{Modifying the rootfs on partition 3}}
\label{fig:part3-noverify-chroot}
\end{figure}
\end{itemize}

\begin{figure}[htbp]
  \centering
    \includegraphics[scale=.25]{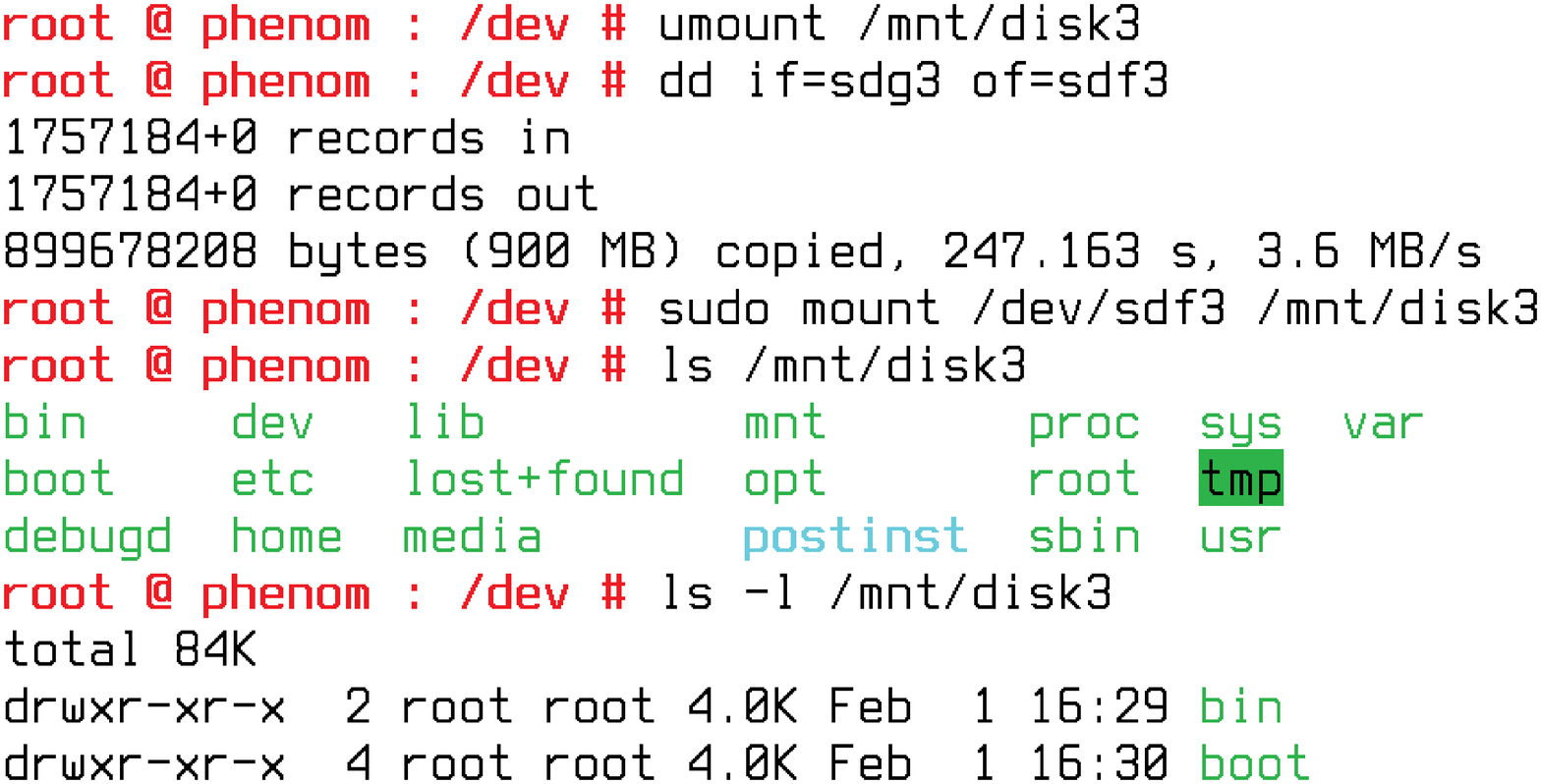}
\caption{\small{Overwriting partition 3 on Chromium-verified boot with partition 3 with no verified boot support}}
\label{fig:verified-overwrite-by-noverify}
\end{figure}

\begin{figure*}[htbp]
  \centering
    \includegraphics[scale=.25]{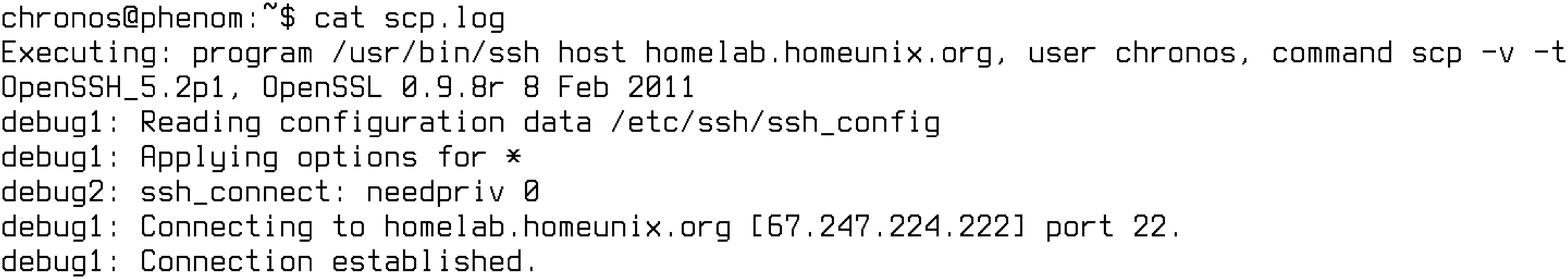}
\caption{\small{Secure copy log of user machine shows the activity of spyware}}
\label{fig:transfer-log}
\end{figure*}
\emph{Replacing the rootfs on verified boot image}: For this phase, the adversary physically acquires the user disk containing Chromium OS with verified boot support. However, the method to tamper the rootfs partition described in the previous section does not work by default on the Chromium OS image with verified boot support. Specifically, partition 3, containing the rootfs, is not mountable, so the rootfs cannot be accessed (Figure~\ref{fig:dmesg-mount-verified}) by default. So, as long as the original rootfs partition and bootloader are intact, the verified boot can detect tampering with the contents of rootfs. But, if the rootfs partition and bootloader are overwritten, the verified boot process is not capable of detecting it. Based on this:
\begin{itemize}
\item The adversary can overwrite partition 3 of verified boot Chromium OS with the tampered rootfs partition created in the previous phase (Figure~\ref{fig:verified-overwrite-by-noverify}).

\item It also overwrites partition 12 of the verified boot Chromium OS with the adversary's partition 12 where the verification is disabled.
 
\item Now, the adversary unmounts the system and boots into Chromium OS and switches to the console mode by pressing (\texttt{Ctrl+Alt+F2}). The adversary uses the phony credential to gain superuser privilege and sets up the following cron job at system startup which will work as the spyware. 

{\tt \small
\begin{verbatim}
* * * * * scp /home/chronos/user/History 
chronos@homelab.homeunix.org:.
\end{verbatim}
}

The spyware will run in the background and copy a user's cached data (in this case, web browsing history) to a machine under the adversary's control. Figure~\ref{fig:attack-environment} gives an overview of machine setups for the exploit. More specifically, this command would Secure Copy (scp) the logged in account user's history file on the user machine to the adversary machine \textit{homelab.homeunix.org} (where the adversary would have already enabled passwordless login) at 1 minute intervals. 
\item Once an unsuspecting user, say Alice, logged in with her Google account on a device containing this tampered rootfs, the cron job would start sending her history (\emph{or any other data as directed by the adversary}) to the adversary's machine, without Alice ever taking notice of it. 

\end{itemize}

Figure~\ref{fig:transfer-log} shows the \texttt{scp} log from the user machine showing the spyware activity. 

\item \emph{Disabling the rootfs verification feature}: An adversary with advanced technical know-how can disable the rootfs verification by physically modifying the partitions without overwriting them. Two advantages of this approach over the previous exploit are that this approach is much faster than the previous exploit of overwriting the rootfs and it is stealthier than the previous one since this exploit preserves the system level user accounts of the victim OS.
For this exploit, an adversary, Eve, does the following:

\begin{itemize}

\item Open victim's rootfs in a hex editor (for e.g. hexcurse),
\item Change the value at \texttt{0x467} from \texttt{FF} to \texttt{00},

\begin{figure*}[htbp]
  \centering
   \includegraphics[scale=0.3]{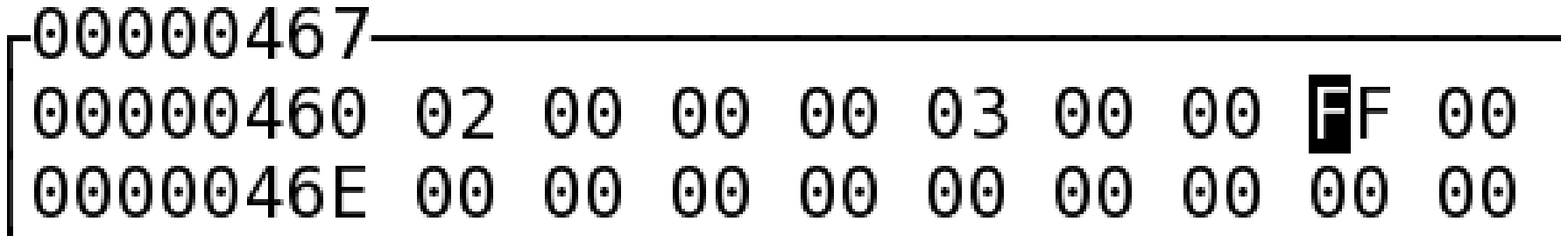}
\caption{\small{Disabling the verification bits}}
\label{fig:hex-attack}
\end{figure*}
\item Save the file,
\item Overwrite the device's bootloader partition with a bootloader partition obtained from another Chromium OS image which doesn't contain any rootfs verification.
\end{itemize}

The resulting image consists of a rootfs which is under Eve's
control, with rootfs verification being disabled while the original
rootfs users continue to exist on it.

Eve can \texttt{chroot} into this rootfs and add additional users and give them
superuser privileges, she can also boot the resulting image and install
spyware on it (as explained in the previous attack).

\end{enumerate}
\section{Mitigation}

Our mitigation technique is centered around detecting modifications
in the bootloader and verification info. When the user, Alice, logs into her Google account, all her files would be decrypted and available
to the spyware on a tampered device. Accordingly, an ideal mitigation technique should work without requiring a login to Alice's Google account. 
\begin{figure*}[htbp]
  \centering
   \includegraphics[scale=0.3]{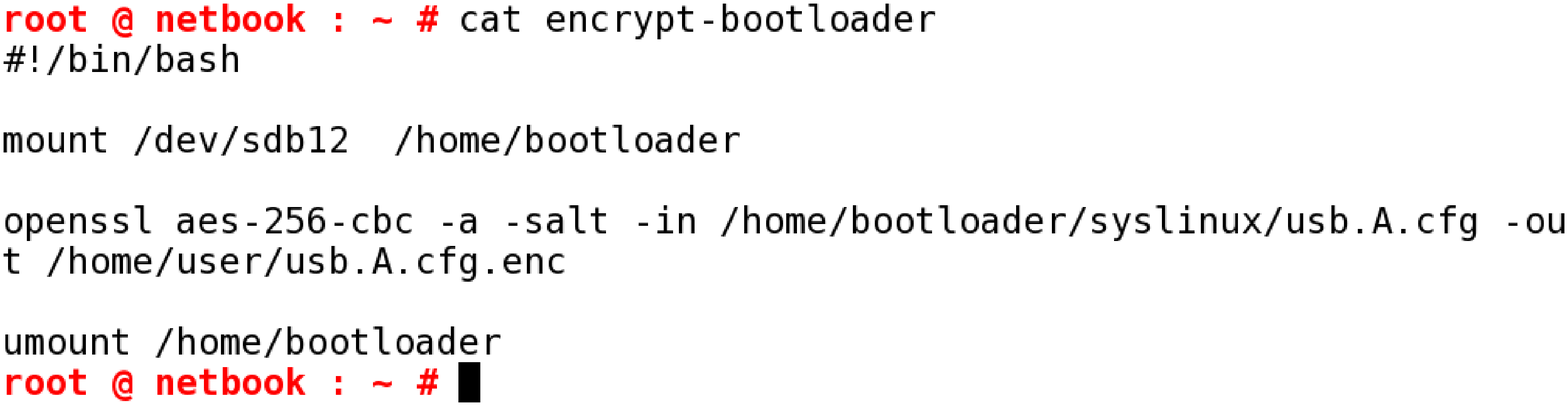}
\caption{\small{Encrypting the contents of bootloader}}
\label{fig:encrypt-bl}
\end{figure*}

Our mitigation techniques work from the system (backend) console without requiring a login to a Google account. Once the image is built, Alice runs the \texttt{encrypt-bootloader} (Figure{\ref{fig:encrypt-bl}}) script which 
creates an encrypted copy of the bootloader configuration file using OpenSSL~\cite{openssl} and stores it in partition 1.
The encryption process uses a password provided by Alice. This password should be ideally different from her Google account password and her rootfs user password.

When the device has been away from Alice's possession for sometime, she shouldn't rule out the possibility of
it being tampered with, and should run the \texttt{verify-integrity} script to check for modifications.
\begin{figure*}[htbp]
  \centering
   \includegraphics[scale=0.3]{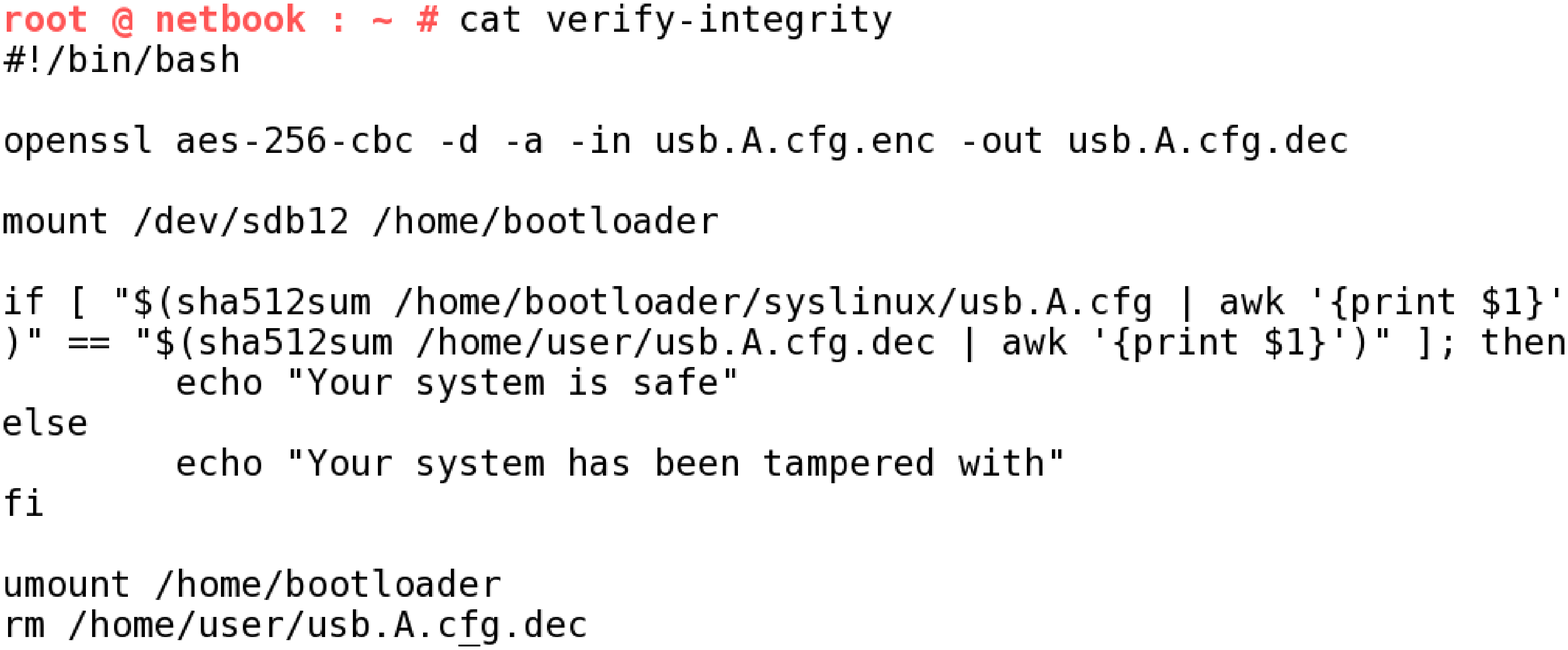}
\caption{\small{Verifying the contents of bootloader}}
\label{fig:verify-bl}
\end{figure*}

The \texttt{verify-integrity} (Figure~\ref{fig:verify-bl}) script does the following: 
\begin{itemize}
\item Decrypt the encrypted bootloader using the password supplied while
encrypting it,
\item Copy the bootloader stored in partition 12 by mounting partition 12
to \texttt{/home/bootloader},
\item Compare the checksum (SHA~\cite{sha1}) of the bootloaders,
\begin{itemize}
\item If the checksums match, the bootloader hasn't been modified and 
the rootfs integrity is verified,
\item If not, then the bootloader has been modified and there is a possible modification of the rootfs.
\end{itemize}

\item Delete the decrypted file and unmount partition 12.
\end{itemize}

\textit{Discussion}: Consider the adversary, Eve, getting access to a device containing these
mitigation scripts. If she performs exploit 1, the users on the rootfs will get
overwritten with Eve's users, and Alice won't be able to log in to the rootfs at all, which is an
indicator that her device has been tampered with. If Eve performs exploit 2, then the original user accounts on the rootfs will be preserved and Alice will still be able to login. Now, when the bootloader is modified, it's checksum will be changed and will be detected
by the \texttt{verify-integrity} script when it's run by Alice. Eve could also insert her own encrypted bootloader in
place of the original file. But this file won't be decrypted using Alice's original password, which will also notify Alice of possible modification.
If Eve deletes the mitigation scripts, then the user would notice the absence of these scripts and conclude that the device
has been tampered with.

\textit{Overhead}: The overhead of our proposed mitigation technique is very low in terms of required time to execute. The \texttt{encrypt-bootloader} script takes $0.1$ sec and the \texttt{verify-integrity} script takes $0.12$ sec to execute on an average. The timing is measured on a netbook with Intel Atom N2600 1.60GHz with 1GB RAM. The overhead of the mitigation technique is negligible and should not hinder a user's experience of fast boot time in a Chromium device.

\section{Conclusion}
In this paper, we have demonstrated a vulnerability in Chromium OS verified boot process that enables a dedicated adversary to launch attacks targeting sensitive user data. Although we have only demonstrated the installation of a spyware to steal the cached user data, in practice a dedicated adversary can also launch severe attacks such as installing a keylogger to steal more sensitive user data such as password, credit card information and so on. Due to the portable usage pattern of Chromium OS devices (netbooks or USB disks), it is not difficult to analyze the practicality of the attack. While we work towards the mitigation of the vulnerability at the source code level, we hope that Chromium OS will go through more rigorous security analysis by the community.


\bibliographystyle{plain}
\bibliography{chrome}


\end{document}